\documentstyle[psfig,floats,preprint,aps]{revtex}
 \tightenlines
 \def\Section#1{}
 
\def\beq{\begin{equation}}
\def\eeq{\end{equation}}
\def\bea{\begin{eqnarray}}
\def\eea{\end{eqnarray}}

\def\D{{\cal{D}}}
\def\Las{{\cal L}}
\begin{document}
\tolerance 50000
\preprint{
\begin{minipage}[t]{1.8in}
\rightline{hep-th/9802014}
\rightline{La Plata-Th 98/03}
\rightline{SISSA 15/98/EP}
\rightline{}
\end{minipage}
}

\draft

\title{Non-Abelian Bosonization and Haldane's Conjecture}

\author{D.C.\ Cabra$^{1}$, P.\ Pujol$^{2}$, C.\ von\ Reichenbach$^{1}$
}
\address{
$^{1}$Departamento de F\'{\i}sica, Universidad Nacional de la Plata,
      C.C.\ 67, (1900) La Plata, Argentina.\\
$^{2}$International School for Advanced Studies,
      Via Beirut 2-4, 34014 Trieste, Italy.\\
}

\date{January 27, 1998}
\maketitle
\begin{abstract}
\begin{center}
\parbox{14cm}{
We study the long wavelength limit of a 
spin $S$ Heisenberg antiferromagnetic chain.
The fermionic Lagrangian obtained corresponds to a 
perturbed level $2S$ $SU(2)$ Wess-Zumino-Witten 
model. This effective theory is then mapped into a compact $U(1)$  boson
interacting with  $Z_{2S}$ parafermions. The analysis of
this effective theory allows us to show that when $S$ is an integer there
is a mass gap to all excitations, whereas this gap vanishes in the
half-odd-integer spin case. This gives a field theory treatment of the 
so-called Haldane's conjecture for arbitrary values of the spin $S$.
}
\end{center}
\end{abstract}

 \vskip1pc]

\Section{Introduction}

In 1983 Haldane conjectured that in the case of integer spin, the spin $S$
quantum Heisenberg Antiferromagnetic (HAF) chain has a unique disordered 
ground state with a finite excitation gap, while the same model has no 
excitation gap when $S$ is a half-odd-integer \cite{Hal}. 
Using a mapping to the non-linear $\sigma$ Model, valid in the large $S$ 
limit, the origin of the difference  has been identified
as being due to an extra topological ``$\Theta$ term" in the effective 
$\sigma$-model Lagrangian for systems with half-integer $S$ \cite{Hal}.
This clearly suggests that the origin of this difference is non-perturbative.
Although Haldane's predictions were 
based on large-$S$ arguments, it is known that this conjecture is
consistent with  the  Bethe Ansatz exact solution
which is available for $S=1/2$ \cite{BA}, and experimental, numerical, and 
theoretical studies
almost confirmed its validity for $S = 1$ \cite{S1}. For higher 
half-odd-integer spin the 
ground state is either degenerate or has a massless excitation \cite{AL} which
suggests but does not prove critical correlations. In \cite{ZS}  massless 
behavior in $S=3/2$ has been tested numerically (see also
\cite{KN}). Recent experimental evidence of the existence of the 
Haldane gap for $S=2$ HAF chains has been found in the study 
of the compound $MnCl_3(bipy)$  \cite{exp}.

Bosonization techniques have been extensively used in the study 
of spin chains (see \cite{LesH} 
for a review on the subject and references therein).
The starting point consists on the mapping of  
the original problem, in terms of 
spin variables, to the problem of fermionic variables with additional 
constraints (which are necessary for the two systems to be equivalent) and 
then 
bosonize the resulting system. In \cite{AH}, the HAF model
was formulated as a certain limit of the Hubbard model. By using a 
bosonization 
technique and a renormalization group analysis,
they found an effective theory for  the low-energy
physics of $S=1/2$ and the limit of large $S$ spin chains. In \cite{Schulz} 
another approach was used which relies on the fact that the
spin $S$ HAF chain can be represented as $2S$ spin $1/2$ chains
for large ferromagnetic coupling.
In a recent paper \cite{IM} the issue of implementing explicitly the
above mentioned additional constraints within the path-integral approach was 
taken up, 
and a reconfirmation of Haldane's conjecture was obtained for large
values of the spin $S$, by using non-abelian bosonization techniques.

In this Letter we study the effective low energy theory in terms of a 
fermionic 
coset model which corresponds to the level $2S$ $SU(2)$ Wess-Zumino-Witten 
(WZW)
theory \cite{Wi,KZ} following \cite{IM}, and map it into a coupled system of 
a compact $U(1)$ boson and 
$Z_{2S}$-Parafermions (PF). The analysis of the phase diagrams of 
$Z_{2S}$ models together with the knowledge of their operator product
algebra provide us with the elements to
establish the difference between integer and half-integer spin chains. 

We would like to stress that our analysis does not rely on a large $S$ 
approximation, but is valid for all values of $S$.

\Section{Coset Model}

Let us  briefly review the work  of  \cite{IM}: 
The spin $S$ HAF model  can be written in terms of fermionic operators
$C_{\alpha i x}$, with $\alpha =\uparrow, \downarrow; i = 1,...,2S$. $\alpha$,
$i$ and $x$ are spin, color and site indices, respectively.
The spin $S$ operator on one site $x$ is represented by

\beq
\vec{S}_x = C^{\dagger}_{\alpha i x} \frac{\vec{\sigma}_{\alpha \beta}}{2}
C_{\beta i x},
\label{spinop}
\eeq
where $\vec{\sigma}_{\alpha \beta}$ are the Pauli matrices. In order to 
correctly
represent the spin $S$ chain, the physical states must satisfy
\bea
\sum_i && C^{\dagger}_{\alpha i x}  C_{\alpha i x} |phys> = 2S|phys>\nonumber\\
\sum_{ij}&& C^{\dagger}_{\alpha i x} \tau^a_{ij} C_{\alpha j y} |phys> = 0,
\label{a}
\eea
where $\tau^a$ are the $SU(2S)$ generators.
The first constraint imposes the condition that allows only one spin per site,
whereas the second one states that the physical states must be color singlets.

The Heisenberg Hamiltonian
\beq
H = \sum_{<xy>} \vec{S}_x . \vec{S}_y
\eeq
can be expressed in terms of the fermionic operators  introduced above
as
\beq
H = - \frac{1}{2} \sum_{<xy>} C^{\dagger}_{\alpha i x} C_{\alpha j y} 
C^{\dagger}_{\beta j y}
C_{\beta i x} + constant.
\eeq
which has a local $SU(2S)\times U(1)$ invariance. This quartic interaction 
can be rewritten by introducing an auxiliary field $B$ as
\beq
H = \frac{1}{2} \sum_{<xy>} ( B^{ij}_{xy} C^{\dagger}_{\alpha i x}
C_{\alpha jy} + h.c. + \bar{B}^{ji}_{yx} B^{ij}_{xy}).
\eeq
In the mean field approximation $B$ is a constant $2S \times 2S$ matrix, and
$H$ can be diagonalized. To obtain an effective low energy theory
we keep
the operators $C(k)$ with $k$ near the Fermi surface $\pm \pi /2a$. We can
then write
\beq
\frac{1}{\sqrt{a}} C_{\alpha i x} = e^{\frac{i \pi}{2a}x} \Psi_{R\alpha i} (x)
+ e^{-\frac{i \pi}{2a}x} \Psi_{L\alpha i} (x),
\label{lel}
\eeq
where $\frac{1}{\sqrt{a}}$ appears for dimensional reasons and 
$ \Psi_{R,L\alpha i} (x)$ are slowly varying on the lattice scale.

We also expand the field $B$ as
\beq
B_{xy} = B_0 e^{aV_{xy}}\simeq B_0 (1 + a V_{xy}),
\label{b}
\eeq
and define $A_1 \equiv \frac{1}{2} (V_{xy} - V^{\dagger}_{xy})$ and
$R_{xy} \equiv \frac{1}{2} (V_{xy} + V^{\dagger}_{xy})$. When substituted
back into
the Hamiltonian, the expansion (\ref{b}) leads to a quadratic integral in
$R_{xy}$ which can be performed to give

\bea
H = B_0 \left(-i \Psi^{\dagger}_{Ri}(\delta_{ij} \partial_x + A^1_{ij}) 
\Psi_{Rj} + \right. \ \ \ \ \ \ \ \ \ \ \ \ \ \ \ \ \ \ \ \   \nonumber \\
\left. i \Psi^{\dagger}_{Li}(\delta_{ij} \partial_x + A^1_{ij})
\Psi_{Lj}\right)
+ \frac{1}{4} \left(\Psi^{\dagger}_{Li}\Psi_{Rj} - \Psi^{\dagger}_{Ri}
\Psi_{Lj}\right)^2,
\label{c}
\eea

The effective Lagrangian is obtained by
introducing a Lagrange multiplier $A_0$ in the Lie algebra of $U(2S)$, 
together with the use of the identity
\beq
\delta[\bar{\Psi_i}\Psi_j] = \lim_{\lambda_2 \rightarrow \infty} e^{- \lambda_2
\int d^2x~(\bar{\Psi}_i\Psi_j)^2},
\eeq
in order to implement  the constraints (\ref{a}) (see \cite{IM} for details).
The effective Lagrangian then reads

\beq
{\it L} = \bar{\Psi}  \gamma^{\mu}i D_{\mu} \Psi - \lambda_1 (i \bar{\Psi}_i
\gamma_5 \Psi_j )^2 - \lambda_2 (i\bar{\Psi}_i \Psi_j)^2,
\label{a1}
\eeq
where $D_{\mu} = \partial_{\mu} - i a_{\mu} + B_{\mu}$, and we have decomposed,
for later convenience, the field $A_{\mu}$ into a $U(1)$ field $a_{\mu}$
and a $SU(2S)$ field $B_{\mu}$.

The Lagrangian can be rewritten as

\bea
{\it L}& =& \bar{\Psi}_{i\alpha} \gamma^{\mu} i (\partial_{\mu} -i a_{\mu}
\delta_{ij}\delta_{\alpha\beta} + B_{\mu}^{ij} \delta_{\alpha \beta}) 
\Psi_{j \beta} +\nonumber\\ 
&+&4 (\lambda_1+ \lambda_2) \vec{J}_R. \vec{J}_L + (\lambda_1 + \lambda_2)
j_Rj_L +\nonumber\\
&-&(\lambda_{1}-\lambda_2)(\Psi^{\dagger}_{Ri\alpha} \Psi_{Lj \alpha}
\Psi_{Rj\beta}^{\dagger} \Psi_{Li \beta} + h.c.),
\label{a3}
\eea
where $\vec{J}_{R,L} = \Psi^{\dagger}_{R,L i\alpha} \frac{\vec{\sigma}_{\alpha 
\beta}}{2}\Psi_{R,L i \beta}$, $j_{R,L} = i \Psi^{\dagger}_{R,L i \alpha} 
\Psi_{R,L i \alpha}$
are $SU(2)_{2S}$ and $U(1)$ currents 
respectively.

The first term in the Lagrangian  (\ref{a3}) corresponds to the fermionic coset 
version of the level $2S$ $SU(2)$ WZW 
theory \cite{NS} as was already observed in
\cite{IM}. In this context the original spin operator corresponds to the
fundamental field of the WZW model.
The third term can be 
absorbed by a redefinition of the $U(1)$ gauge field $a_{\mu}$.

Thus, we have to deal with the second and last terms in (\ref{a3}) which 
can be expressed as

\beq
\Delta\Las = (\lambda_{1}-\lambda_{2}) (g_{\alpha\beta}g_{\beta\alpha} + 
h.c.) +
4 (\lambda_1+ \lambda_2) \vec{J}_R. \vec{J}_L 
\label{b1}
\eeq
where $g\equiv \Psi_{Ri}^{\dagger} \Psi_{Li }$ is 
the spin 1/2 primary field of the $SU(2)_{2S}$ WZW theory,
$\Phi^{(1/2)}$, with
conformal dimensions  $ h = \bar{h} = 3 / (8(S+1))$. The first term in 
(\ref{b1})
corresponds to the spin $1$ affine primary $\Phi^{(1)}$ with conformal 
dimensions
$h=\bar h=1/(S+1)$, so we can write

\beq
\Delta\Las = -4 ~(\lambda_1-\lambda_2)~ tr~\Phi^{(1)}
+ 4 (\lambda_1+ \lambda_2) \vec{J}_R. \vec{J}_L 
\label{b3}
\eeq

The $S = \frac{1}{2}$ case is simpler, as has been discussed in \cite{AH},
\cite{IM}, since affine (Kac-Moody) selection 
rules forbid the appearance of the relevant
operator  $\Phi^{(1)}$. We then have an effective 
massless theory  in accordance with Haldane's
predictions. The second term in (\ref{b3}) is
marginally irrelevant since $\lambda_1+ \lambda_2$
is positive, and gives the well-known logarithmic corrections
to correlators.

For higher spins, we have to consider the interaction term
(\ref{b3})  and we also have to include all other terms which will be
radiatively generated. We then need the
operator product expansion (OPE) coefficients among the different 
components of $\Phi^{(1)}$ which have been computed in \cite{FZ}.
The OPE coefficients are non-vanishing
iff the so called ``Fusion Rules"  are non-vanishing. In the level $k$ $SU(2)$ 
WZW theory they are given by \cite{Verlinde}
\beq
\Phi^{(j)}_{m,\bar m}\times  \Phi^{(j')}_{m',\bar m'} =
\sum_{n=\vert j-j'\vert}^{min(j+j',k-j-j')}  \Phi^{(n)}_{m+m',\bar m + \bar m'}
\label{fr}
\eeq

We will use that \cite{FZ,GQ}
\beq
SU(2)_k \equiv Z_k \otimes U(1)
\label{equiv}
\eeq
in the sense that the Hilbert spaces of
the two theories coincide. We will exploit this equivalence 
to derive an effective low energy action for the spin $S$ HAF chain.
Indeed, it was shown in  \cite{FZ}  that the primary fields of the
$SU(2)_k$ WZW theory are related to the
primaries of the $Z_k$-parafermion theory and the $U(1)$ vertex operators.
They are connected by
the relation

\beq
\Phi^{(j)}_{m, \bar{m}}(z, \bar{z}) = \phi^{(2j)}_{2m,2\bar{m}}(z, \bar{z})
:e^{\frac{i}{\sqrt{2S}}(m\varphi (z)+ \bar{m}\bar{\varphi}(\bar{z}))}:,
\eeq
where the $\Phi$ fields are the invariant fields of the $SU(2)_k$ WZW theory, 
the $\phi$ fields are the
$Z_{k}$ PF primaries and $\varphi$ and $\bar{\varphi}$ are 
the holomorphic and antiholomorphic components of a compact massless
free boson field.
In the same way,  the currents are related as
\bea
J_R^+ (z)&=& (2S)^{1/2} \psi_1 (z) :\exp\left({i \over \sqrt{2S}} 
\varphi(z)\right): ~,~ \nonumber \\
J_R^0 (z) &=& (2S)^{1/2} \partial_z \varphi (z) 
\label{curr}
\eea
where $\psi_1$ is the first parafermionic field. (A similar 
relation holds for the left-handed currents).

Using this equivalence we can express 
the relevant perturbation term  (\ref{b3}) as

\bea
\Delta\Las =- 4 (\lambda_1-\lambda_2) \left(\phi^{(2)}_{0,0} + 
\phi^{(2)}_{2,-2}
:e^{\frac{i}{\sqrt{2S}}(\varphi (z)-
 \bar{\varphi}(\bar{z}))}: \right.
\nonumber \\
\left. +\phi^{(2)}_{-2,2} :e^{-\frac{i}{\sqrt{2S}}(\varphi 
(z)- \bar{\varphi}(\bar{z}))}:\right)\nonumber \\
 +4S(\lambda_1+ \lambda_2) \left( \psi_1 \bar \psi_1^{\dagger} 
:e^{\frac{i}{\sqrt{2S}}(\varphi (z)- \bar{\varphi}(\bar{z}))}:
 + h.c.\right),
 \label{h1}
\eea
where we absorbed the derivative part of the $U(1)$ field coming 
from (\ref{curr}) into a redefinition of the constant
in front of the unperturbed Lagrangian. 
The first term corresponds to the first ``thermal" field of the PF theory,
$\phi^{(2)}_{0,0}=\epsilon_1$, with conformal dimensions
$h=\bar h=1/(1+S)$, while the second and  
third  terms correspond to the $p=2$ disorder
operator in the PF theory, $\phi^{(2)}_{2,-2}= \mu_2$ and its adjoint
$\phi^{(2)}_{-2,2} = \mu_2^{\dagger}$ with
dimensions $d_2 = \bar{d}_2 = (S-1)/(2S(S+1))$. 

Including all the operators  which are radiatively
generated we  get three ``families" of perturbations:

\noindent 1) The thermal operator $\epsilon_1$ and all the members of its 
sub-algebra (higher thermal operators). As shown in \cite{Fa},
the $Z_{2S}$ PF theory perturbed by $\epsilon_1$ flows into a massive regime,
irrespectively of the sign of the coupling. 
Assuming that, as for the $Z_2$ case, due to the sign of the coupling 
$\lambda_1 - \lambda_2$ in (\ref{h1}) the theory is driven into a low 
temperature 
ordered phase, we have that vacuum expectation values (v.e.v.'s) of  disorder 
operators $\mu_j$, vanish for $ j \neq 2S~mod(2S) $ as 
well as v.e.v.'s of the parafermionic fields $\langle \psi_{k} 
\bar{\psi}_k^{\dagger} 
\rangle = 0$, for  $ 2k \neq 2S~mod(2S) $. 

\noindent 2) The family of the disorder operator (with the corresponding 
vertex operators) given by:
$$
\sum_{k=1}^{[S]} \mu_{2k} :e^{\frac{ik}{\sqrt{2S}}(\varphi (z)-
 \bar{\varphi}(\bar{z}))}: + h.c.
$$
where $[S]$ means integer part of $S$. Note that $\mu_{2S}$ 
corresponds to the identity operator 
and then for $S$ integer, this family contains the single vertex operator
\beq
:e^{\frac{iS}{\sqrt{2S}}(\varphi (z)-
 \bar{\varphi}(\bar{z}))}: + h.c.
 \label{rel}
\eeq

\noindent 3) The family of the parafermionic fields:
$$
\sum_{k=1}^{2S} \psi_{k}(z) {\bar \psi}_k^{\dagger}(\bar z)
:e^{\frac{ik}{\sqrt{2S}}(\varphi (z)-  \bar{\varphi}(\bar{z}))}: + h.c.
$$

Note that the system is invariant under the 
extended $Z_{2S} \times \tilde{Z}_{2S}$ transformation
\bea
\mu_p \rightarrow w^{p(m-n)} \mu_p \nonumber ~;~ 
\psi_p {\bar \psi}_p^{\dagger} \rightarrow w^{2p(m-n)} 
\psi_p {\bar \psi}_p^{\dagger}~;~ \nonumber \\
\varphi \rightarrow \varphi - {\sqrt{2} \pi  m \over \sqrt{S}} ~;~
\bar{\varphi} \rightarrow \bar{\varphi} - {\sqrt{2} \pi  n \over \sqrt{S}} 
\label{transf}
\eea
with $w=\exp(\pi i  /S)$ and $m,n \in Z$. 

Since the parafermionic sector is massive, the 
effective theory for large scales  can be obtained by integrating out these 
degrees of freedom. Since
this massive sector is driven into the phase where (\ref{transf}) is unbroken,
we can obtain the most general effective  action
for the  remaining $U(1)$ field, where only the vertex operators invariant 
under (\ref{transf}) will be allowed to appear. We get
\bea
Z_{eff}= \int \D \phi~\exp \left(-\int K_S (\partial_{\mu} \phi)^2 + \right.
\nonumber \\
\left. \alpha_S  \int \cos({\frac{2S}{\sqrt{2S}}
(\varphi - \bar{\varphi})}) + \cdots\right),
\label{zeffsem}
\eea
for $S$ half-integer, and 
\bea
Z_{eff}= \int \D \phi~\exp \left(-\int K_S (\partial_{\mu} \phi)^2 + 
\right. \nonumber \\
\left. \beta_S
\int \cos({S \over \sqrt{2S}} (\varphi - \bar{\varphi})) + 
\right. \nonumber \\
\left. \alpha_S \int  \cos( {2S \over \sqrt{2S}} (\varphi - \bar{\varphi})) 
+ \cdots \right),
\label{zeffint}
\eea
for $S$ integer. Here the dots simply mean higher powers of the
perturbing vertex operators and
$K_S$ is an effective constant arising from the OPE 
of vertex and parafermionic operators in the process
of integration of the massive degrees
of freedom. We immediately notice that for integer $S$ there is an extra
vertex operator coming from (\ref{rel}) which is not present for half-integer 
$S$, and as we will see, 
this difference between integer and half-integer effective actions is crucial.

Using (\ref{spinop}) and (\ref{lel}) we can write the continuum expression of 
the original spin operator $\vec S(x)$ as
\beq
\vec S(x)=\vec J_R+\vec J_L+(-1)^x tr(\vec \sigma/2 (\Phi^{(1/2)}+\Phi^{(1/2) 
\dagger})).
\eeq
One way to see whether the system is gapped or 
not is to study the  behavior of the 
spin-spin correlation function at large scales. Since
our original $SU(2)$ WZW model is perturbed, 
correlation functions of the fundamental
field will contain supplementary operators coming 
from the OPE between $\Phi^{(1/2)}$
and the perturbing fields. With the help of the 
fusion rules (\ref{fr}) it is easy to see that, for example, 
the effective alternating z-component of the spin operator 
containing the scalar field will be given by:
\beq
\sum_{k \leq 2S,~k~odd} a_k ~\mu_{k} :e^{\frac{ik}{2\sqrt{2S}}(\varphi
(z)-
 \bar{\varphi}(\bar{z}))}: + h.c. ,
\label{spin}
\eeq
where only odd $k$ fields appear in the sum.
Let us consider now separately the case of half-integer and integer $S$: For
$S$ half-integer, the operator $\Phi^{(S)}$ is 
present in (\ref{spin}), and we can easily
check that, (since $\mu_{2S}$ corresponds 
to the identity), this operator is simply given 
by 
$$
e^{\frac{iS}{\sqrt{2S}}(\varphi - \bar{\varphi})}+ h.c.
$$ 
The other operators in the series contain parafermionic disorder
operators whose correlators will decay exponentially to zero at large scales. 
Thus, considering only the Gaussian part of
(\ref{zeffsem}), we can show that the spin correlation functions at large 
scales behave like:
\bea
<S_z(x) S_z(y)> \sim (-1)^{(x-y)}  \vert x-y \vert^{-2SK_S} \nonumber \\
<S_+(x) S_-(y)> \sim (-1)^{(x-y)} \vert x-y \vert^{-1/(2SK_S)}
\eea
The fact that the $SU(2)$ symmetry is unbroken 
at all scales fixes then the value of $K_S$ to be
\beq
K_S = 1/(2S)
\label{k}
\eeq
For this value of $K_S$ one can show that the perturbing operator in 
(\ref{zeffsem}) is marginally irrelevant.

We conclude then that the large scale behavior of  half-integer 
spin chains is given by the level 1 $SU(2)$ WZW model with logarithmic
corrections as for the spin $1/2$ chain. An 
interesting extension of this analysis
is to add an arbitrary dimerization to the chain. 
In the context of the WZW approach
this corresponds to the addition of the field $tr\ \Phi^{(1/2)} + h.c.\propto 
\mu_{1} :e^{\frac{i}{2\sqrt{2S}}(\varphi (z)-
 \bar{\varphi}(\bar{z}))}: + h.c.$ 
in the perturbing terms \cite{LesH}. 
Reproducing the 
same arguments as above, we see that this term has the effect of 
making the operator $\cos({\frac{S}{\sqrt{2S}}
(\varphi - \bar{\varphi})})$ appear in the effective 
action (\ref{zeffsem}). Using (\ref{k}), 
we see that now this operator is highly
relevant and will so produce a gap in 
the excitation spectrum unless a fine tuning of 
the parameters is performed.

Let us consider now  integer spins $S$. Since the series (\ref{spin}) for the 
effective spin operator contains only half-integer spins $j$ (odd $k$'s), 
all the operators in the series will contain non-trivial parafermionic 
operators. Then all the terms in the spin-spin correlation function will
decay exponentially
to zero with the distance indicating the presence of a gap in the excitation
spectrum, thus confirming Haldane's 
conjecture. This is our main result.

A possible modification consists in the addition of 
dimerization to the system (which again 
corresponds to the inclusion of  the field $tr\ \Phi^{(1/2)}$
as a perturbation). Then, also integer spin $j$ terms (even $k$'s) will be 
generated (as indicated by the fusion rules (\ref{fr}))
in the effective spin operator and in particular the vertex operator 
(\ref{rel}) which is the only candidate
for a power law decay of the spin-spin correlators. However, as for the 
(dimerized)  half-integer spin case,
the presence in (\ref{zeffint}) of the (relevant) first vertex operator prevent
the system to be massless (this situation is similar to the one encountered in
\cite{Schu2}, in the context of abelian bosonization),
indicating again that a fine tuning of the dimerization
parameter has to be performed to get a massless regime.

We have presented the continuum limit of the spin-$S$ antiferromagnetic
Heisenberg chain in terms of a parafermion Conformal Field Theory interacting 
with a compact $U(1)$ boson. After integrating out the massive parafermions
we have shown that HAF chains with half-odd-integer spins have
a massless spectrum while those with integer spin have a gap to all
excitations, in complete accordance with Haldane's conjecture.

It has been shown  by using a mapping to the sigma model
that a dimerized spin $S$
chain should have $2S+1$ massless points 
in the dimerization parameter space \cite{AH}.
This result was shown to be valid for large $S$ 
and a detailed treatment of dimerization
within the present approach could help  to extend this result for small
values of the spin.

\smallskip

We are grateful to F.\ Alcaraz, P.\ Dorey, V.S.\ Dotsenko,  A.\ Lugo, 
V.A.\ Fateev, A.\ Honecker, M.\ Picco, V.\ Rittenberg for useful discussions.
D.C.C.\  thanks CONICET  and Fundaci\'on Antorchas for financial \
support. C.v.R. is supported by CONICET, Argentina.

\end{document}